\documentstyle[prd,aps]{revtex}

\begin{document}
\preprint{SUSSEX-AST 96/7-4, astro-ph/9607096}
\draft

%
% Remove this and closure after abstract in electronic submission
%
\input epsf

\twocolumn[\hsize\textwidth\columnwidth\hsize\csname 
@twocolumnfalse\endcsname

\title{Accurate determination of inflationary perturbations}
\author{Ian J. Grivell and Andrew R. Liddle}
\address{Astronomy Centre, University of Sussex, Falmer, Brighton BN1 
9QH, United Kingdom.}
\date{\today}
\maketitle
\begin{abstract}
We use a numerical code for accurate computation of the amplitude of 
linear density perturbations and gravitational waves generated by 
single-field inflation models to study the accuracy of existing analytic 
results based on the slow-roll approximation. We use our code to calculate 
the coefficient of an expansion about the exact analytic result for 
power-law inflation; this generates a fitting function which can be applied 
to all inflationary models to obtain extremely accurate results. In the 
appropriate limit our results confirm the Stewart--Lyth analytic 
second-order calculation, and we find that their results are very accurate 
for inflationary models favoured by current observational constraints.
\end{abstract}
\pacs{PACS numbers: 98.80.Cq \hspace*{4cm} SUSSEX-AST 96/7-4, 
astro-ph/9607096}

\vskip2pc]

%%%%%%%%%%%%%%%%%%%%%%%%%%%%%%%%%%%%%%%%%%%%%%%%%%%%%%%%%%%%%%%%%%%%%%%%
\section{Introduction}

In anticipation of a near-future launch of a satellite capable of 
measuring microwave background anisotropies to an accuracy of a few percent 
or better across a wide range of angular scales \cite{SAT}, attention has 
recently been directed towards obtaining highly accurate predictions of the 
anisotropies for given cosmological models. It is now possible for fast and 
extremely accurate calculations, good to a percent or so, of the radiation 
anisotropy power spectrum to be made \cite{HSSW+SZ}, given the spectrum of 
density perturbations in the universe.

A very promising theory for the generation of such perturbations is 
cosmological inflation \cite{INFL}, wherein the universe experiences a 
period of accelerated expansion in its very early stages. In order to take 
advantage of the great observational accuracy anticipated, it is desirable 
to make as accurate calculations as possible of the perturbations which 
inflation produces. Such accurate calculations are the focus of this paper.

When density perturbations from inflation were first discussed 
\cite{PERT,BST}, the result was typically quoted along the lines of
\begin{equation}
\left. \frac{\delta \rho}{\rho} \right|_{{\rm hor}} \sim \; \left.
	\frac{H^2}{|\dot{\phi}|}\right|_{{\rm hor}} \,.
\end{equation}
where the right hand side is evaluated as the scale crosses outside the 
Hubble radius during inflation, and the left hand side as it crosses back 
inside after inflation. Here $H$ is the Hubble parameter and $\phi$ is 
the scalar field driving inflation. That the result was given only 
approximately was through a combination of the calculations not being 
completely accurate, and the lack of a precise definition of the left hand 
side of this equation, usually described as the fractional density 
perturbation when a scale enters the horizon. Fairly quickly, however, a 
precise definition of this quantity was made and the coefficient computed 
\cite{BST,lyth85}. In more modern notation, defined later in this paper, the 
result is
\begin{equation}
\label{lowestsr}
{\cal P}_{{\cal R}}^{1/2}(k) \simeq \left. \frac{H^2}{2\pi|\dot{\phi}|}
	\right|_{k=aH} \,.
\end{equation}
The right hand side is evaluated as before, but now the left hand side, 
giving the spectrum ${\cal P}$ of the curvature perturbation ${\cal R}$, 
holds at any time when the scale is much larger than the Hubble radius, the 
curvature perturbation ${\cal R}$ being constant in that limit.
The uncertainty in the calculation was then due to the fact that its 
derivation depended on the inflationary slow-roll approximation; for some 
models this can be extraordinarily good but in many the associated error 
would be expected to be order of ten percent or greater.

The next development was the presentation of a precise calculation in the 
case of power-law inflation, by Lyth and Stewart \cite{LS}. The analogous 
but simpler exact calculation for the gravitational wave amplitude in this 
model was made much earlier by Abbott and Wise \cite{AW}. The precise 
calculation allowed a direct evaluation of the uncertainty induced in these 
models by use of the slow-roll approximation. Stewart and Lyth then went on 
to use this exact result to analytically compute the next-order slow-roll 
correction to the standard formula \cite{SL}. In principle, one would like 
to carry out the expansion about any power-law inflation model, but in 
practice analytic considerations required a restriction to an expansion 
about power-law inflation models which were themselves near the de Sitter 
limit. One should therefore regard that calculation as a slow-roll expansion 
rather than an expansion about the power-law inflation solution.

A slightly different kind of calculation, which we shall not further 
discuss, can be done in the case where the potential takes the form 
$V(\phi) = V_0 \pm m^2 \phi^2/2$, in the limit where $V_0$ dominates. For 
the inverted harmonic oscillator (natural inflation) this was carried out by 
Stewart and Lyth \cite{SL} and for positive mass (hybrid inflation) an 
analysis was performed by Garc\'{\i}a-Bellido and Wands \cite{GBW}. There a 
much more accurate calculation can be made than the slow-roll one, even if 
the mass is quite large, but again this calculation is restricted to this 
particular class of models.

For most models of inflation, the Stewart--Lyth calculation should give 
extremely accurate predictions. In this paper, we use a numerical code to 
test their accuracy against exact results, finding that the agreement is 
extremely good. We then use our code as the basis for an even more 
accurate scheme, using it to carry out an expansion about any power-law 
inflation solution, even those far from the de Sitter limit. This gives a 
highly accurate fitting function for the amplitude in any inflation model, 
by regarding them as expansions about the `nearest' power-law inflation 
model.

\section{The Density Perturbation Calculation}

\subsection{Evolution of perturbations}

Standard calculations of the density perturbation amplitude from 
inflationary models rely on two separate types of assumption, linear 
perturbation theory and the slow-roll approximation for the inflationary 
evolution of the background space-time. The first of these approximations is 
considerably better than the second, since we know from COBE that the 
amplitude of perturbations is about one part in $10^5$, and we shall use it 
throughout.

The slow-roll approximation is typically much less good, because in order 
for inflation to end the slow-roll conditions must break. While for some 
inflationary models it is extremely good, it is perfectly possible for it to 
fail by ten percent or more in others, including some still permitted by 
present observational constraints. Hence our aim in this paper to make 
calculations which do not depend on the slow-roll approximation, and also to 
provide more accurate approximations for the density perturbations based on 
a version of the slow-roll expansion.

Before proceeding to the calculation, let us highlight that we are limiting 
discussion to inflation driven by a single, canonically normalized, scalar 
field in Einstein gravity. While this looks like quite a restrictive 
assumption, is is actually fairly general, in the sense that almost any 
extended theory of gravity, such as a higher-order or scalar--tensor theory, 
can be rewritten as Einstein gravity via a conformal transformation. 
Further, models with more than one scalar field often effectively only have 
a single degree of freedom, with fluctuations transverse to the evolving 
field having their fluctuations suppressed by a large effective mass; this 
situation occurs, for example, when extended inflation is conformally 
transformed to give power-law inflation \cite{extinf}, and also in the 
hybrid inflation model \cite{hybrid}. However, our calculations will not be 
valid in any theory in which fluctuations in more than one field are 
important, and it is possible to construct situations where this is the case 
\cite{twoperts}.

Our analysis is based on the formalism devised by Stewart and Lyth \cite{SL} 
to carry out a second-order analytic calculation. We shall not reproduce the 
derivation of the equations here, instead simply reproducing the important 
ones.\footnote{An expanded discussion of the Stewart--Lyth calculation is 
given in Ref.~\cite{LLKCBA}.} With the usual notation 
of $a$ for scale factor, $H$ for Hubble parameter, $\phi$ for the 
inflationary scalar field and overdot as derivative with respect to cosmic 
time, they introduce a new quantity $z$ defined by
\begin{equation}
\label{z}
z\equiv \frac{a\dot{\phi}}{H} \,.
\end{equation}
The quantity one desires to calculate is the curvature perturbation ${\cal 
R}$, defined as in Ref.~\cite{SL}. It is convenient to define a related 
quantity $u$, defined by
\begin{equation}
\label{uR}
u = - z {\cal R} \,,
\end{equation}
which is a gauge invariant potential \cite{MFB}.

A Fourier expansion of $u$ into comoving modes $u_{{\bf k}}$ is carried out, 
and these can be shown to obey the remarkably simple equation 
\cite{Muk,MFB,SL}
\begin{equation}
\label{ufield}
\frac{d^2 u_{{\bf k}}}{d\tau^2} +\left( k^2 - \frac{1}{z} 
	\frac{d^2 z}{d\tau^2} \right) u_{{\bf k}} =0 \,,
\end{equation}
where $\tau$ is the conformal time defined by $d\tau = dt/a(t)$ and $k = 
|{\bf k}|$ is the modulus of the wavenumber. Defining the spectrum ${\cal 
P}_{{\cal R}}(k)$ of the curvature perturbation in the standard way as  
\begin{equation}
\langle {\cal R}_{{\bf k_1}} {\cal R}^*_{{\bf k_2}} \rangle =
	\frac{2\pi^2}{k^3} {\cal P}_{{\cal R}} \delta^{3} \,
	({\bf k_1}-{\bf k_2}) \,,
\end{equation}
yields
\begin{equation}
\label{pspec}
{\cal P}_{{\cal R}}^{1/2}(k) = \sqrt{\frac{k^3}{2\pi^2}} \, 
	\left| \frac{u_{{\bf k}}}{z} \right| \,.
\end{equation}
which statistical isotropy permits to depend only on the modulus $k$. From 
here on, we'll label modes by their modulus $k$ rather than their full 
wavenumber.

During inflation, a comoving scale evolves from well inside the Hubble 
radius to well outside it. Canonical quantization of $u$ demands that in the 
initial state, where the expansion can be ignored, it approaches the 
standard flat space result \cite{SL}
\begin{equation}
\label{short}
u_k (\tau) \rightarrow \frac{1}{\sqrt{2 k}} \; e^{-ik\tau}\,.
\end{equation}
In the opposite regime, where $k$ can be ignored in Eq.~(\ref{ufield}), the 
equation can be integrated directly to give
\begin{equation}
\label{long}
u_k \propto z \,,
\end{equation}
implying that the spectrum ${\cal P}_{{\cal R}}(k)$ approaches a constant in 
this regime. 

Notice that the entire quantum part of this calculation is in the 
normalization at large $k$, which is the Minkowski space limit. Once that 
has been determined, then in a Heisenberg picture the state vector is time 
independent and the mode functions $u_k$ obey the classical 
equations of motion which allow us to evolve them. The full theory of 
quantum fields in curved space-time is not required.

The crucial question, of course, is the value of the proportionality 
constant in Eq.~(\ref{long}). We have seen that in both the short and long 
wavelength regimes the evolution of amplitude is independent of the way in 
which the universe expands. The crucial feature which determines the final 
amplitude is therefore the way in which the universe evolves as the mode 
switches from one asymptotic regime to the other. This requires a modelling 
of the relevant inflationary epoch.

\subsection{Inflation and the slow-roll approximation}

Given a specific inflationary potential $V(\phi)$ for the scalar field 
$\phi$, there is a well posed problem for the evolution of the background 
space time. The relevant behaviour that we want to extract is that of 
$(d^2 z/d\tau^2)/z$, as required in Eq.~(\ref{ufield}).

The most compact way of describing inflationary solutions is the 
Hamilton--Jacobi formalism \cite{HJ}, where one adopts the scalar field 
itself as the time variable and in which the solution is described via 
$H(\phi)$. The equations of motion read
\begin{eqnarray}
\left( \frac{dH}{d\phi} \right)^2 - \frac{12\pi}{m_{{\rm Pl}}^2} 
	H^2 (\phi) & = & - \frac{32\pi^2}{m_{{\rm Pl}}^4} V(\phi) \,, \\
-\frac{m_{{\rm Pl}}^2}{4\pi} \frac{dH}{d\phi} & = & \dot{\phi} 
	\; \left( \; = \frac{d \phi}{a\, d \tau} \; \right) \,.
\end{eqnarray}
If required, the potential corresponding to any solution $H(\phi)$ is easily 
obtained from the first of these. 

{}From this fundamental quantity, we can introduce a series of functions 
containing higher and higher derivatives of $H(\phi)$ via\footnote{The 
functions $\eta$ and $\xi$ are equivalent to $\beta_1$ and $\beta_2$; we 
give them a special name as they crop up frequently below. The first 
function, $\epsilon$, doesn't fit into the pattern of the rest.}
\begin{eqnarray}
\epsilon(\phi) & \equiv & \frac{m_{{\rm Pl}}^2}{4 \pi} \left( \frac{H'}{H}
	\right)^2 \,, \\
\eta(\phi) & \equiv & \frac{m_{{\rm Pl}}^2}{4 \pi} \frac{H''}{H} \,,\\
\xi(\phi) & \equiv & \frac{m_{{\rm Pl}}^2}{4 \pi} \left(
	\frac{H' H'''}{H^2} \right)^{1/2} \,, \\
\beta_n(\phi) & \equiv & \frac{m_{{\rm Pl}}^2}{4\pi} \left( 
	\frac{ \left( H' \right)^{n-1} H^{(n+1)}}{H^n}
	\right)^{1/n} \quad \left( n \geq 3 \right) \,,
\end{eqnarray}
where prime is a derivative with respect to $\phi$ and `$(n+1)$' symbolizes 
the taking of $n+1$ derivatives with respect to $\phi$. These functions are 
known as the slow-roll parameters, introduced by Liddle, Parsons and Barrow 
\cite{LPB}. They form the basis for the slow-roll expansion, which allows 
arbitrarily accurate solutions to the dynamical equations governing 
inflation to be obtained. For our purposes, they are useful because we can 
replace the general function $H(\phi)$ by the values of $H$ and these 
slow-roll parameters at a single value of $\phi$ --- this is equivalent 
information because, modulo questions of convergence, the full function 
$H(\phi)$ can be reconstructed via a Taylor expansion about this point. 
Because we are interested primarily in the behaviour as a scale $k$ crosses 
the Hubble radius, we shall choose the value of $\phi$ when $k = aH$. Notice 
that the calculation we carry out is for a single value of $k$; even with 
the same inflationary potential, the values of these expansion coefficients 
change if we change the scale $k$, since that corresponds to moving to a 
different part of the potential.

As we mentioned at the start of the Section, the inflationary input into the 
perturbation equation is that it determines $(d^2 z/d\tau^2)/z$ in 
Eq.~(\ref{ufield}). This quantity can be rewritten in terms of the slow-roll 
parameters, as
\begin{equation}
\label{zderiv}
\frac{1}{z} \frac{d^2z}{d\tau^2} = 2a^2 H^2 \left[ 1+\epsilon -
	\frac{3}{2} \eta + \epsilon^2 -2 \epsilon \eta + \frac{1}{2} 
	\eta^2 + \frac{1}{2} \xi^2 \right] \,.
\end{equation}
Although this looks like it might be the start of an expansion, it is 
actually exact.

\subsection{Exact solution for power-law inflation}

We begin by rederiving the exact solution describing power-law inflation, 
following Lyth and Stewart \cite{LS,SL}. The evolution of the background 
space-time is given by
\begin{equation}
\label{pliH}
H(\phi) = H_0 \exp \left( \sqrt{\frac{4\pi}{p}} \, 
	\frac{\phi-\phi_0}{m_{{\rm Pl}}} \right) \,,
\end{equation}
where $H_0$ and $p >1$ are constants. The scalar field is translated so that 
$\phi = \phi_0$ at the time when the scale $k$ we are interested in obeys $k 
= aH$. This case is the simplest because all the slow-roll parameters are 
constant, and in fact they are all equal to $1/p$. This means that 
Eq.~(\ref{zderiv}) takes on a particularly simple form.

We also require an expression for the conformal time, which for power-law 
inflation can be derived using a trick of integrating by parts
\begin{equation}
\tau \equiv \int \frac{dt}{a(t)} = \int \frac{da}{a^2 H} = 
	-\frac{1}{aH} + \int \frac{\epsilon \, da}{a^2 H} \,,
\end{equation}
which for constant $\epsilon$ implies
\begin{equation}
\tau = -\frac{1}{aH} \, \frac{1}{1-\epsilon} \,.
\end{equation}
	
With these results, Eq.~(\ref{ufield}) for the perturbations reduces to a 
Bessel equation
\begin{equation}
\label{bessel}
\left[ \frac{d^2}{d\tau^2} +k^2 -\frac{(\nu^2 - 1/4)}{\tau^2} 
	\right] u_k =0 \,,
\end{equation}
where
\begin{equation}
\nu \equiv \frac{3}{2} + \frac{\epsilon}{1-\epsilon} \,,
\end{equation}
is a constant. The solution with the correct short-scale behaviour, shown in 
Eq.~(\ref{short}), is
\begin{equation}
u_k(\tau) = \frac{\sqrt{\pi}}{2} e^{i(\nu +1/2)\pi/2} 
	(-\tau)^{1/2} H_{\nu}^{(1)} (-k\tau) \,,
\end{equation}
where $H_{\nu}^{(1)}$ is the Hankel function of the first kind of order
$\nu$. 

The result we desire is the asymptotic form of the solution; taking $k/aH 
\rightarrow 0$ gives the asymptotic form
\begin{equation}
u_k \rightarrow e^{i(\nu -1/2)\pi /2} 2^{\nu -3/2} 
	\frac{\Gamma (\nu)}{\Gamma (3/2)} \frac{1}{\sqrt{2k}} 
	(-k \tau )^{-\nu +1/2} \,,
\end{equation}
where $\Gamma$ is the usual gamma function. On substitution into the 
expression for the power spectrum, Eq.~(\ref{pspec}), this gives 
\begin{equation}
\label{scalaramp}
{\cal P}_{{\cal R}}^{1/2}(k) =2^{\nu -3/2} \frac{\Gamma(\nu)}{\Gamma(3/2)} 
	(\nu-1/2)^{1/2 - \nu} \frac{2}{m_{{\rm Pl}}^2} \left. 
	\frac{H^2}{|H'|} \right|_{k=aH} \,.
\end{equation}
It should be stressed that, despite appearances, this equation does not give 
the value of the perturbation as it crosses the Hubble radius. Instead, it 
gives the asymptotic value as $k/aH \rightarrow 0$, rewritten in terms of 
the values which quantities had at Hubble radius crossing.

In the limit of small $\epsilon$, this approaches the standard result
\begin{equation}
\label{specsmall}
{\cal P}_{{\cal R}}^{1/2}(k) \simeq \left. \frac{H^2}{2\pi|\dot{\phi}|}
	\right|_{k=aH} = \left. \frac{1}{\sqrt{\pi \epsilon}} 
	\, \frac{H}{m_{{\rm Pl}}} \right|_{k=aH}\,.
\end{equation}
Note however that the amplitude diverges in that limit.

\subsection{Strategy for a general calculation}

The aim is to use this exact solution to estimate the amplitude expected in 
any inflationary model, by relating the expansion behaviour crucial for the 
perturbation generation on a given scale $k$ to that of a suitable power-law 
inflation model. Power-law inflation in effect gives a two parameter set of 
solutions, the input parameters being $H_0$ and $\epsilon=1/p$. In a general 
inflation model $H$ and $\epsilon$ vary in some way with time, and take on 
some particular values when $k = aH$ (at $\phi = \phi_0$). We can then 
choose a reference power-law inflation model which matches these values; 
this guarantees that the amplitude and slope of $H(\phi)$ about $\phi_0$ are 
the same in the reference power-law model as in the true model. The 
perturbation amplitude is then computed in the reference model. Since the 
final answer is primarily determined by the behaviour when $k \sim aH$, this 
generates the appropriate approximate answer, provided the true $H(\phi)$ 
respects this linear approximation sufficiently accurately across the 
relevant scales. This reasoning leads to the standard result, 
Eq.~(\ref{specsmall}). 

Notice that the calculation is carried out for a single scale $k$; to 
generate the complete spectrum, then at each $k$ there is a different 
reference power-law inflation model because in general both $H$ and 
$\epsilon$ evolve in a way different to power-law inflation.

If one follows the logic of the preceding paragraphs, then it comes as a 
surprise that the standard result quoted is the small $\epsilon$ result, 
Eq.~(\ref{specsmall}), rather than the exact power-law inflation result 
Eq.~(\ref{scalaramp}). One reason for this is that the level of accuracy of 
including the $\epsilon$-dependent prefactor has usually not been required. 
However, a more important reason is that the prefactor in 
Eq.~(\ref{scalaramp}), which depends only fairly weakly on $\epsilon$, is 
not general enough; there are terms connected to $\eta$ (i.e.~from the 
second derivative of $H$) which give comparable corrections. For a more 
accurate result than Eq.~(\ref{specsmall}), these need to be taken into 
account.

Stewart and Lyth \cite{SL} used the power-law inflation solution as the 
basis for an analytic expansion intended to hold for all inflation models; 
in an approximation where $\epsilon$ and $\eta$ are treated as negligibly 
varying they obtained
\begin{equation}
\label{sl}
{\cal P}_{{\cal R}}^{1/2}(k) \simeq  \left[1 - (2C+1) \epsilon + C \eta
	\right] \left. \frac{H^2}{2\pi|\dot{\phi}|} \right|_{k=aH} \,,
\end{equation}
where $C = - 2 + \ln 2 + \gamma \simeq -0.73$ is a numerical constant 
($\gamma$ being the Euler constant). The slow-roll parameters are also 
evaluated at $k = aH$. The approximation scheme used to derive this result 
relies on assuming $\epsilon$ is small as well as $\eta$, and therefore 
the full strength of the exact power-law inflation result is not being used.

We consider more general circumstances, by carrying out a general
expansion about any power-law solution. Any general background model
can be specified by giving the function $H(\phi)$, or equivalently the
value \mbox{$H_{0}=H(\phi_{0})$} and the values of all the slow-roll
parameters at horizon crossing for the mode of interest. We can then
express $H(\phi)$ in terms of the corresponding power-law expression
(with the same values of $H_{0}$ and $\epsilon$) and a series of
correction terms which depend on the higher--order slow-roll
parameters:
\begin{eqnarray}
\label{H_exp}
\frac{H(\phi)}{H_{0}} & = & \exp \left[ \sqrt{4 \pi \epsilon}\, 
	\frac{\phi - \phi_{0}} {m_{\rm Pl}} \right] +  \\
 & & \sum_{n=2}^{\infty} \frac{1}{n!}
	\left[ \left( \frac{\beta_{n-1}}{\epsilon} \right)^{n-1} - 1 
	\right] \left[ \sqrt{4 \pi \epsilon}\, 
	\frac{\phi - \phi_{0}}{m_{\rm Pl}} \right]^{n} \nonumber \,.
\end{eqnarray}

The quantity we are interested in, ${\cal P}_{{\cal R}}^{1/2}$, is a
functional of $H(\phi)$ and is therefore a function of the slow-roll
parameters in the same form as they appear in the expression for $H$
above:
\begin{equation}
{\cal P}_{{\cal R}}^{1/2} = {\cal P}_{{\cal R}}^{1/2}(\epsilon,
\sigma_{2}, \sigma_{3}, \ldots)
\end{equation}
where we have defined the new set of parameters
\begin{equation}
\sigma_{n} = \left[ \left( \frac{\beta_{n-1}}{\epsilon} \right)^{n-1}
	- 1 \right]\epsilon^{n/2}\,.
\end{equation}
These parameters give a measure of the `distance' of the true model from
the reference power-law model; they are all to be evaluated at
\mbox{$k=aH$} so they are just numbers, not functions of $\phi$. One
should think of these as measuring the amount by which the
derivatives of $H$ at Hubble-radius-crossing fail to match those
expected of the power-law solution given its amplitude and first
derivative. For models close to power-law inflation these parameters
will be small and in such cases continuity demands that ${\cal
P}_{{\cal R}}^{1/2}$ receives a small correction from the power-law
result.\footnote{For this to be strictly true, it is necessary that the
functional ${\cal P}_{{\cal R}}^{1/2}$ depends only on a limited range
of values of $H(\phi)$, which we have seen is the case here.} This
allows us to write the spectrum as an expansion in powers of small
parameters; to first-order in \mbox{$\sigma_{2},\sigma_{3},\ldots$}
the expansion is
\begin{equation}
\label{p_exp}
{\cal P}_{{\cal R}}^{1/2}(k) = [1 + A(\epsilon)\sigma_{2} +
	B(\epsilon)\sigma_{3} + \cdots]
	\tilde{{\cal P}}_{{\cal R}}^{1/2}(k) 
\end{equation}
where $\tilde{{\cal P}}_{{\cal R}}^{1/2}(k)$ is the exact power-law
solution Eq.~(\ref{scalaramp}) for the appropriate $\epsilon$ and
where everything on the right hand side is evaluated at
\mbox{$k=aH$}. Here \mbox{$A(\epsilon),B(\epsilon),\ldots$} are
calculable functions giving, at each value of $\epsilon$, the
coefficient in the expansion of each $\sigma_n$. In the limit of small
$\epsilon$, $A(\epsilon)$ must approach the value predicted by the
Stewart--Lyth calculation
\begin{equation}
A(\epsilon)\rightarrow C {\rm ~~as~~ } \epsilon \rightarrow 0 
\end{equation}
where $C=-2 + \ln 2 + \gamma \simeq -0.73$ is a numerical constant,
$\gamma$ being the Euler constant.

In principle, to completely specify the background model we require
values for an infinite number of slow-roll parameters, so even
restricting the expansion Eq.~(\ref{p_exp}) to first-order in each
$\sigma_n$ still leaves us with an infinite number of terms. However,
as we have pointed out the perturbation spectrum depends on $H(\phi)$
only for some interval of $\phi$ around $\phi_0$ and so in a Taylor
series expansion of $H$ around $\phi_0$ we expect the higher-order
terms (involving higher-order parameters $\sigma_n$) to be
successively less significant. We therefore try a truncated form of
the expansion Eq.~(\ref{p_exp}), keeping only the first few terms.

Unfortunately, we have not been able to determine $A(\epsilon)$
analytically. Instead therefore we resort to a numerical computation.

\section{Numerical solution of the perturbation equations}

\begin{figure}
\centering 
\leavevmode\epsfysize=6cm \epsfbox{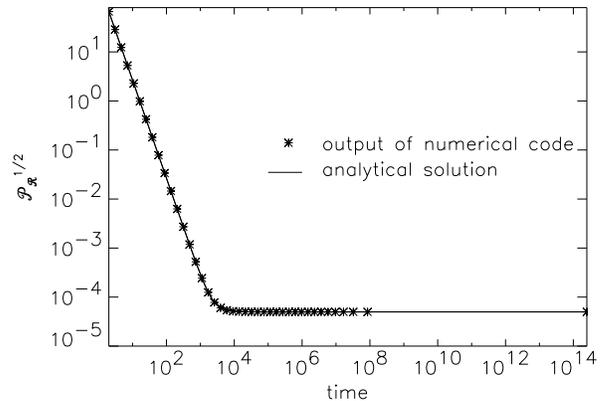}\\ 
\caption[plitest]{\label{plitest} A comparison of numerical and analytical 
solutions for perturbations generated by power-law inflation.} 
\end{figure}

The perturbation equation is solved numerically. As an initial test, we 
check that the code can reproduce the analytical power-law inflation 
solution to high accuracy. In fact the asymptotic power spectrum
output by the code differs from the analytical result by less than one
part in $10^5$. Fig.~\ref{plitest} shows a comparison 
between the output of our numerical code and the exact solution. 

It can be seen from Fig.~\ref{plitest} that before horizon crossing
the perturbation amplitude decreases with time. At sufficiently early
times the amplitude of the perturbation is large with respect to the
background and the linear theory breaks down \cite{MFB}. We therefore
do not expect the large initial values of the amplitude displayed in
Fig.~\ref{plitest} to be accurate. However, before horizon crossing
the amplitude is fixed by the standard QFT normalization,
Eq.~(\ref{short}), and is independent of the earlier behaviour, so since
the perturbation becomes small before horizon crossing our
results for the final amplitude using linear theory will be valid.

Easther \cite{E} has recently constructed another inflationary model for 
which the perturbation equations can be solved exactly; we have also 
successfully tested that our code can reproduce this solution. As it lacks 
many of the nice properties of power-law inflation, it does not appear 
promising for using as the basis of an expansion.

Having thus confirmed the accuracy of the code, we can compare the
numerical results with those obtained from analytical approximations
based on the slow-roll approximation. As input to the code we use
$H(\phi)$ as given by Eq.~(\ref{H_exp}) and solve the perturbation
equation for a range of values of $\epsilon$ and $\eta$ (or
equivalently $\sigma_2$) for the mode which crosses the horizon at
\mbox{$\phi=\phi_0$}. All higher-order $\sigma_n$ are set to zero. 

\begin{figure}
\centering 
\leavevmode\epsfysize=10cm \epsfbox{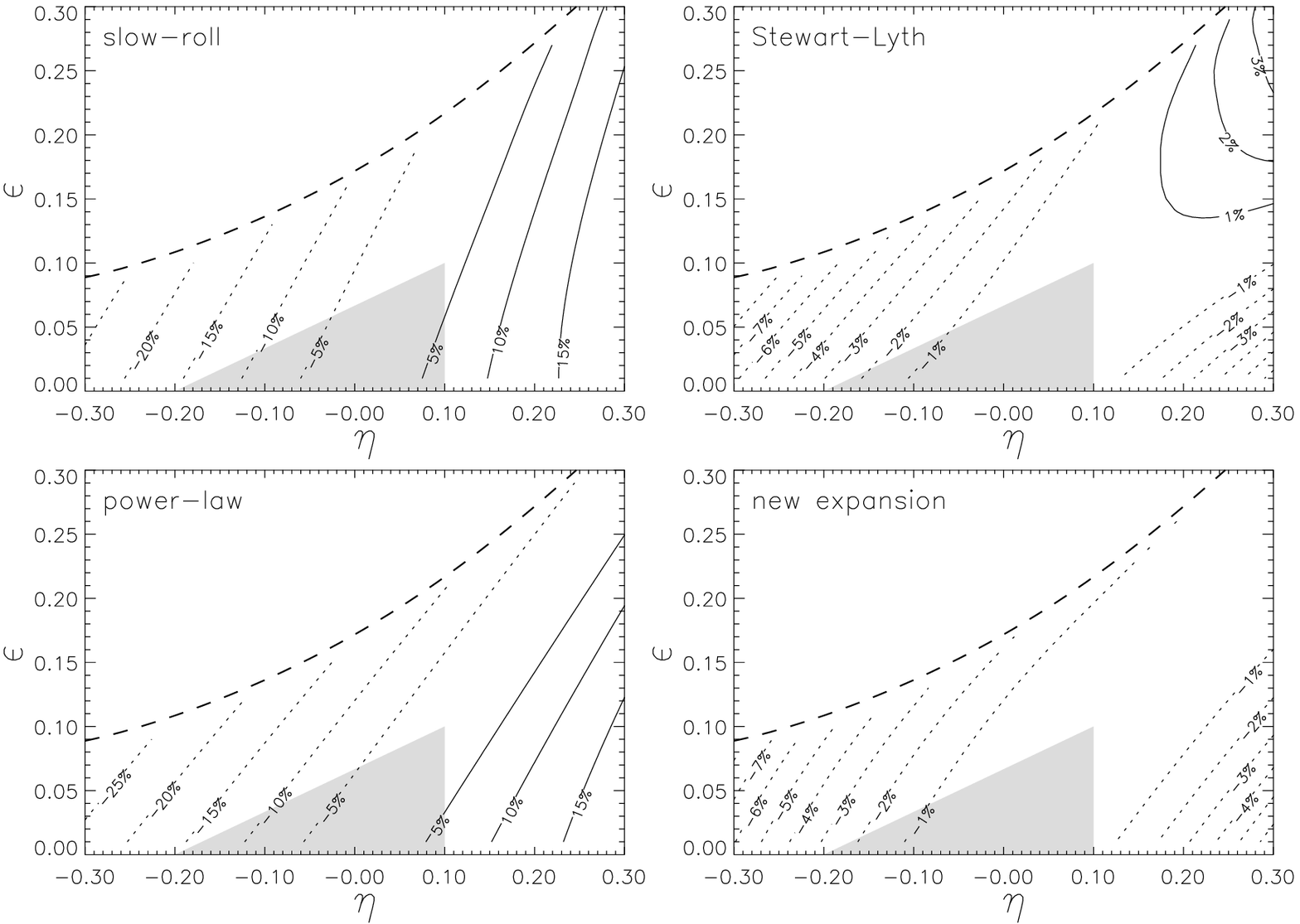}\\ 
\caption[contours]{\label{contours} Contours indicating the relative 
differences between the numerical results and each of the approximate 
expressions for the amplitude of scalar perturbations. Dashed contours 
indicate an underestimate of the true result and solid ones an 
overestimate, by the percentage indicated. Above the thick dashed line the 
asymptotic regime is not reached before inflation ends, so no results are 
available. The shaded region indicates the parameters favoured by current 
observations.}
\end{figure}

In Fig.~\ref{contours} we display the relative differences between the
numerical results and each of the four different approximations we have
considered here: the lowest-order slow-roll expression,
Eq.~(\ref{lowestsr}), the Stewart--Lyth calculation, Eq.~(\ref{sl}), the
exact power-law solution for the corresponding value of $\epsilon$,
Eq.~(\ref{scalaramp}), and the expansion about the power-law solution
given by Eq.~(\ref{p_exp}). The region above the thick dashed curve in these
contour plots has been excluded, because for these values of the
parameters inflation ends soon after horizon crossing and the
perturbation amplitude has not attained its asymptotic value. The
shaded triangle in these plots represents the region in the 
$\epsilon$--$\eta$ plane favoured by current observations of 
large-scale structure and microwave background anisotropies \cite{LLetc}, 
allowing for the uncertainty in cosmological parameters.

The final panel of Fig.~\ref{contours} gives a comparison of the numerical
results with the expansion Eq.~(\ref{p_exp}) and as such requires that
we know the coefficient $A(\epsilon)$. We have used the numerical code
to evaluate this coefficient and the next-order coefficient
$B(\epsilon)$ and we find that for \mbox{$\epsilon \leq 0.3$} they are
well fitted by the linear expressions
\begin{eqnarray}
A(\epsilon) & \simeq & -0.730 - 0.850 \epsilon \\
B(\epsilon) & \simeq & 0.010 + 0.480 \epsilon \,. \label{b}
\end{eqnarray}
As anticipated, we find the contribution from $\sigma_3$ to be
significantly smaller than that from $\sigma_2$.

Before we compare the numerical results with the approximations it is
worth considering how representative our model is at each value of
$\epsilon$ and $\eta$. In other words, how reasonable is it to
neglect the effect of higher-order parameters? In order to get some
idea of the impact of higher-order terms on the results, we have
repeated the numerical calculations for the values of $\epsilon$
and $\eta$ shown in Fig.~\ref{contours}, using a different expansion for $H$ 
as the input by instead expanding its logarithm:
\begin{eqnarray}
\ln \frac{H(\phi)}{H_0} &= &\sqrt{4 \pi \epsilon} \left( 
	\frac{\phi - \phi_0}{m_{\rm Pl}} \right) + 2\pi \sigma_2 
	\left(\frac {\phi - \phi_0}{m_{\rm Pl}} \right)^2 \\
& & + \frac{(4 \pi)^{3/2}}{6} (\sigma_3 - 3\sigma_2
	\epsilon^ {1/2}) \left( \frac {\phi - \phi_0} {m_{\rm Pl}} 
	\right)^3 + \cdots \nonumber 
\end{eqnarray}
but keeping only the first two terms. In this case we have specific
non-zero values for all of the higher-order parameters in terms of 
$\epsilon$ and $\sigma_2$, with
$\sigma_3$ for example given by
\begin{equation}
\sigma_3 = 3 \sigma_2 \epsilon^{1/2} \,.
\end{equation}
We therefore have numerical results for two different slices through
the parameter space. Comparing the results for the two slices
point-by-point in the \mbox{$\epsilon - \eta$} plane we find that
within the region shown in the contour plots of Fig.~\ref{contours}
the results differ by at most $2\%$. When Eq.~(\ref{b}) is used to
calculate the first-order correction arising from the contribution of
$\sigma_3$, the relative difference between the two sets of
results is found to be below $0.5\%$ within this region. Of course it
will always be possible to construct models where the higher-order
terms are more significant than this, but in most cases the
contributions arising from the parameters beyond $\sigma_3$ will be
negligible. 

\section{Gravitational Waves}

We have also carried out calculations for the somewhat simpler case of 
gravitational waves, where the relevant equation, analogous to 
Eq.~(\ref{ufield}), is \cite{Grish,AW,SL}
\begin{equation}
\label{vfield}
\frac{d^2 v_{{\bf k}}}{d\tau^2} +\left( k^2 - \frac{1}{a} 
	\frac{d^2 a}{d\tau^2} \right) v_{{\bf k}} =0 \,.
\end{equation}
The corresponding power spectrum ${\cal P}_{{\rm g}}^{1/2}$ is expanded in 
the same manner as for scalar perturbations 
\begin{equation}
\label{g_exp}
{\cal P}_{{\rm g}}^{1/2}(k) = [1 + A_{{\rm g}}(\epsilon)\sigma_{2} +
	B_{{\rm g}}(\epsilon)\sigma_{3} + \cdots]
	\tilde{{\cal P}}_{{\rm g}}^{1/2}(k)\,, 
\end{equation}
where once again $\tilde{{\cal P}}_{{\rm g}}^{1/2}$ is the exact power-law
inflation result and the coefficients are determined from numerical
calculations to be
\begin{eqnarray}
A_{{\rm g}} (\epsilon) &\simeq &0.16\epsilon + 2.8\epsilon^2 \\
B_{{\rm g}}(\epsilon) &\simeq &-1.1\epsilon^2\,.
\end{eqnarray}
The absence of a constant term for $A_{{\rm g}}(\epsilon)$ is in agreement
with the Stewart--Lyth result for the gravitational wave spectrum.
The differences between this approximation and the numerical results
are shown in Fig.~\ref{gw}, along with those for the other approximate
expressions. Notice that in this case the power-law inflation result does 
better than the Stewart--Lyth result; this is because for gravitational 
waves, unlike density perturbations, the latter is simply the small 
parameter expansion of the former. Note also that the results for 
gravitational waves are much more accurate than those for density 
perturbations.

\section{Discussion}

In Fig.~\ref{contours} the results based on the slow-roll
approximation and on the exact power-law expression are compared with the
numerical results for the case of scalar perturbations. The lowest-order 
slow-roll result is seen to perform well only for very small values of the 
slow-roll parameters, with errors in excess of 10\% within the 
observationally favoured region. In contrast, the Stewart--Lyth result fares 
surprisingly well. The power-law expression, although exact along the line 
\mbox{$\epsilon = \eta$}, rapidly decreases in accuracy away from this line. 
By construction, our new expansion about this exact solution shows the best 
agreement with the numerical results across the full range of parameters, 
though the improvement only sets in a considerable distance from the 
slow-roll limit. At small $\epsilon$ and $\eta$ the performance of our 
expansion is comparable with the Stewart--Lyth calculation. These results 
show that the assumption \mbox{$\epsilon \ll 1$}, required for the 
Stewart--Lyth result to be applicable, doesn't break down until well outside 
the region constrained by observations. It seems then that for reasonable
inflationary models the Stewart--Lyth calculation works extremely well.

\begin{figure}
\centering 
\leavevmode\epsfysize=10cm \epsfbox{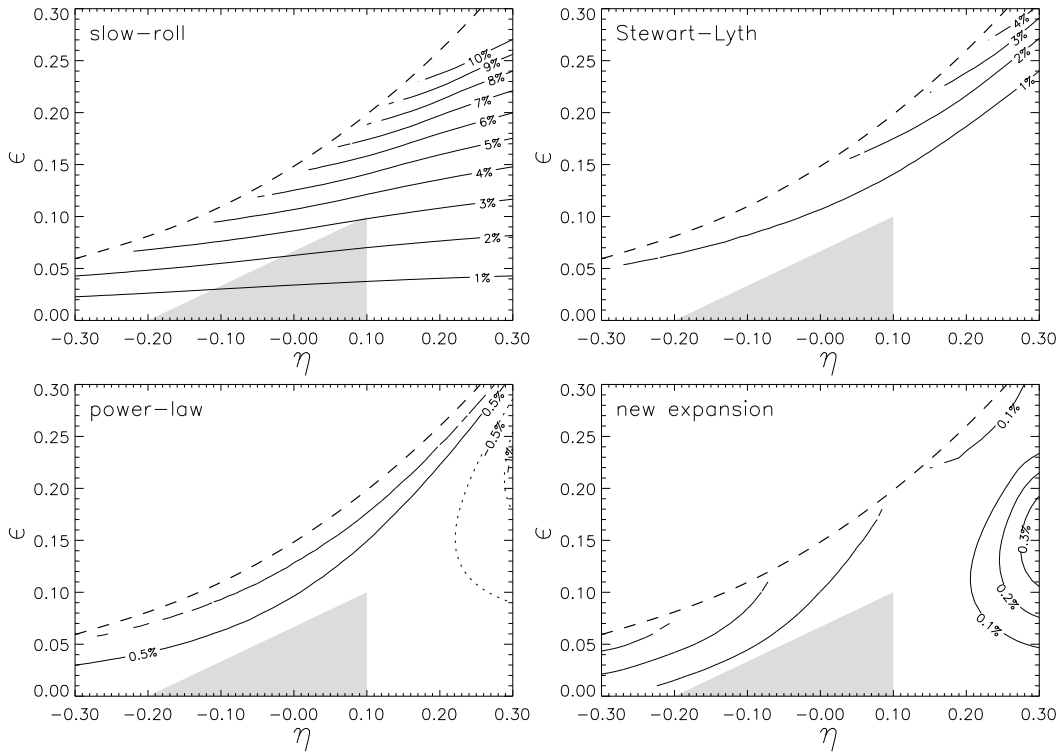}\\ 
\caption[gw]{\label{gw} As Fig.~\ref{contours} but for gravitational 
waves.}
\end{figure}

Fig.~\ref{gw} shows the comparisons between the numerical results
and each of the approximations for the spectrum of gravitational
waves. In this case the exact power-law expression is found to be
extremely accurate. There is considerable improvement in expanding
about this exact solution, but the high level of accuracy achieved is
not likely to ever be required.

As a check on the applicability of these results to general
inflationary models, we have numerically evaluated the perturbation
amplitude for a model with a steep polynomial potential, for the mode 
crossing the horizon 60 \mbox{$e$-foldings} before the end of inflation. In
this calculation the inputs to the numerical code are the potential
and the initial conditions for the scalar field, the latter taken to
be given by the inflationary attractor \cite{LPB}. 
We took the potential
\begin{equation}
V(\phi) = V_0 \left(\frac{\phi}{m_{\rm Pl}} \right)^{\alpha} \quad ;
	\quad \alpha=10 \,.
\end{equation}
in order to be not too near the slow-roll limit; such potentials were 
discussed in Ref.~\cite{LSha}. The slow-roll parameters take the values
\begin{equation}
\epsilon = 0.04 \quad ; \quad \eta = 0.03 \,.
\end{equation}
The comparisons between the approximations and the numerical result show a 
relative error
\begin{eqnarray}
\mbox{lowest order slow-roll:~~} & 5 \times 10^{-3}, \nonumber \\
\mbox{Stewart--Lyth:~~} & 5 \times 10^{-4}, \nonumber \\
\mbox{PLI:~~} & 6 \times 10^{-3}, \nonumber \\
\mbox{new expansion:~~} & 2 \times 10^{-4}.
\end{eqnarray}
Note that for this case all approximation schemes fall within a 1\% error.

To recap, we have used the exact power-law result as the basis for an 
expansion
which gives the perturbation spectrum very accurately for inflationary
models which are `close' to power-law models, in the sense that the
parameters $\sigma_n$ are small. This expansion outperforms the
Stewart--Lyth result for large $\epsilon$, but within the range of
values for the slow-roll parameters allowed by cosmological
observations we have shown that the Stewart--Lyth result is remarkably 
accurate. Our results indicate that the errors in both ours and the 
Stewart--Lyth calculation are due mostly to neglecting terms of order 
$\sigma_2^2$, with the contributions from higher-order parameters being 
relatively insignificant. We stress that these conclusions are only valid if 
the perturbation spectrum is sensitive only to the behaviour of $H(\phi)$ 
close to horizon crossing, but within the single scalar field paradigm we 
have been discussing we expect this to be the norm and the result
calculated for a specific potential is consistent with this.

%%%%%%%%%%%%%%%%%%%%%%%%%%%%%%%%%%%%%%%%%%%%%%%%%%%%%%%%%%%%%%%%%%%%%%%%
\section*{Acknowledgments}

The authors are supported by the Royal Society. We thank Ed Copeland, Jim 
Lidsey and David Lyth for discussions on the analytical second-order 
calculation and Juan Garc\'{\i}a-Bellido, Ewan Stewart and David Wands for 
helpful comments on this work. We acknowledge use of the Starlink computer 
system at the University of Sussex. 

%%%%%%%%%%%%%%%%%%%%%%%%%%%%%%%%%%%%%%%%%%%%%%%%%%%%%%%%%%%%%%%%%%%%%%%%

\clearpage
\widetext

\narrowtext

\end{document}